\documentclass[aps, pre, superscriptaddress, twocolumn]{revtex4}
\usepackage{bm}
\usepackage{amsmath}
\usepackage{amsfonts}
\usepackage{amsthm}
\usepackage{xcolor}
\usepackage[pdftex]{graphicx}
\usepackage[utf8]{inputenc}

\begin{document}

\title{Single-particle velocity distributions of collisionless,\\ steady-state plasmas must follow Superstatistics}

\author{Sergio Davis}
\email{sergio.davis@cchen.cl}

\affiliation{Comisión Chilena de Energía Nuclear, Casilla 188-D, Santiago, Chile}
\affiliation{Departamento de F\'isica, Facultad de Ciencias Exactas, Universidad Andres Bello. Sazi\'e 2212, piso 7, 8370136, Santiago, Chile.}

\author{Gonzalo Avaria}
\affiliation{Comisión Chilena de Energía Nuclear, Casilla 188-D, Santiago, Chile}
\affiliation{Departamento de F\'isica, Facultad de Ciencias Exactas, Universidad Andres Bello. Sazi\'e 2212, piso 7, 8370136, Santiago, Chile.}

\author{Biswajit Bora}
\affiliation{Comisión Chilena de Energía Nuclear, Casilla 188-D, Santiago, Chile}
\affiliation{Departamento de F\'isica, Facultad de Ciencias Exactas, Universidad Andres Bello. Sazi\'e 2212, piso 7, 8370136, Santiago, Chile.}

\author{Jalaj Jain}
\affiliation{Comisión Chilena de Energía Nuclear, Casilla 188-D, Santiago, Chile}

\author{José Moreno}
\affiliation{Comisión Chilena de Energía Nuclear, Casilla 188-D, Santiago, Chile}
\affiliation{Departamento de F\'isica, Facultad de Ciencias Exactas, Universidad Andres Bello. Sazi\'e 2212, piso 7, 8370136, Santiago, Chile.}

\author{Cristian Pavez}
\affiliation{Comisión Chilena de Energía Nuclear, Casilla 188-D, Santiago, Chile}
\affiliation{Departamento de F\'isica, Facultad de Ciencias Exactas, Universidad Andres Bello. Sazi\'e 2212, piso 7, 8370136, Santiago, Chile.}

\author{Leopoldo Soto}
\affiliation{Comisión Chilena de Energía Nuclear, Casilla 188-D, Santiago, Chile}
\affiliation{Departamento de F\'isica, Facultad de Ciencias Exactas, Universidad Andres Bello. Sazi\'e 2212, piso 7, 8370136, Santiago, Chile.}

\date{\today}

\begin{abstract}
The correct modelling of velocity distribution functions for particles in steady-state plasmas is a central element in the study of nuclear fusion and also in the
description of space plasmas. In this work, a statistical mechanical formalism for the description of collisionless plasmas in a steady state is presented, based solely
on the application of the rules of probability and not relying on the concept of entropy. Beck and Cohen's superstatistical framework is recovered as a limiting case,
and a ``microscopic'' definition of inverse temperature $\beta$ is given. Non-extensivity is not invoked \emph{a priori} but enters the picture only through the analysis
of correlations between parts of the system.
\end{abstract}

\maketitle

\section{Introduction}

Plasmas in nonequilibrium steady states sometimes follow non-Maxwellian velocity distributions, one of the most common known as Kappa
distributions~\cite{Lima2000,Leubner2002,Livadiotis2017}. Properties of Kappa-distributed plasmas have been extensively
studied~\cite{Gougam2011,Vinas2015,Livadiotis2014,Livadiotis2016}, however their origin is still a matter of debate.

One of the main arguments put forward for the existence of Kappa distributions in plasmas relies on the non-extensive statistics proposed by Constantino Tsallis~\cite{Tsallis2009},
which postulates a generalization of the Boltzmann-Gibbs entropy depending on a non-extensive index $q$. More recently, the formalism known as superstatistics~\cite{Beck2004}
has been gaining attention, as it is able to produce Kappa distributions without extending the entropy: it asserts that non-Maxwellian distributions arise from a superposition
of canonical ensembles at different values of the inverse temperature $\beta = 1/k_B T$. Interestingly, superstatistics as a tool to explain the non-Maxwellian distributions found
in plasmas has only recently been properly acknowledged in the literature~\cite{Ourabah2015}.

In this work we present a first-principles derivation of the single-particle energy distribution for steady-state plasmas, using only the tools of standard statistical mechanics.
We show that, in general, the velocity distribution of a single particle follows the superstatistical formalism and we give an interpretation of the superstatistical (inverse)
temperature distribution in this case. This work opens the way for a direct test of the superstatistical formalism against explicit kinetic simulations, for instance,
based on particle-in-cell methods.

\section{Statistical description of plasma}
\label{sect_stat_plasma}

As it turns out, in most situations only the high-energy tails of the velocity distribution in plasmas of interest (be it laboratory or space plasmas)
reach relativistic speeds. However, without any additional effort we can obtain the fully relativistic results and then take the low energy limit
for non-relativistic cases. Hence we will start with a fully relativistic statistical mechanical description of $N$ charged particles subjected to fixed,
time-dependent electromagnetic fields. The Lagrangian of such a system is given by

\begin{equation}
L = \sum_{i=1}^N \left(-\frac{m_i c^2}{\gamma(\bm v_i)} - q_i V(\bm r_i; t) + q_i \bm v_i \cdot \bm A(\bm r_i; t)\right),
\label{eq_lagrangian}
\end{equation}
where $V(\bm r; t)$ and $\bm A(\bm r; t)$ are the scalar and vector potentials, respectively, and $\gamma(\bm v)$ is the
Lorentz factor, $$\gamma(\bm v) = 1/\sqrt{1-(v/c)^2}.$$ The canonical momentum $\bm p_i$ associated to $\bm r_i$ is given by
\begin{equation}
\bm p_i = \frac{\partial L}{\partial \bm v_i} = \gamma(\bm v_i) m_i \bm{v_i} + q_i\bm A(\bm r_i; t),
\label{eq_momentum}
\end{equation}
and the Hamiltonian is $H = \sum_{i=1}^N(\bm p_i \cdot \bm v_i) - L$. Let us introduce the $3N$-component vectors $\bm R$ and $\bm V$ as notation
shortcuts for $(\bm r_1, \ldots, \bm r_N)$ and $(\bm v_1, \ldots, \bm v_N)$ respectively. Instead of working in the canonical variables $\bm r_i$
and $\bm p_i$ it is convenient to go back to positions and velocities, as in this way we can write an energy function $E(\bm R, \bm V; t)$, which is nothing
but $H(\bm R, \bm P; t)$ rewritten to eliminate $\bm p_i$ using Eq. \ref{eq_momentum},

\begin{align}
E(\bm R, \bm V; t) & = \sum_{i=1}^N \Big(\gamma(\bm v_i)m_i c^2 + q_iV(\bm r_i; t)\Big) \nonumber \\
                   & = \sum_{i=1}^N K_i(\bm v_i) + \Phi(\bm R; t),
\label{eq_energy}
\end{align}
and which is independent of $\bm{A}$. We see that $E$ is simply the sum of kinetic and potential energies, as the terms involving $q_i \bm v_i\cdot \bm A(\bm r_i; t)$ have cancelled out.

The description of this system in terms of statistical mechanics in absence of collisions is given by the Liouville equation, where we have taken the
6$N$-dimensional vectors $(\bm R, \bm V)$ as the available microstates of the system. The Liouville equation can written in our case in the form~\cite{Ichimaru1992},

\begin{equation}
\left[\frac{\partial}{\partial t} + \sum_{i=1}^N\Big(\bm v_i\cdot \frac{\partial}{\partial \bm r_i} + \bm a_i\cdot \frac{\partial}{\partial \bm v_i}\Big)\right]P(\bm R, \bm V|t) = 0,
\label{eq_continuity}
\end{equation}
where $P=P(\bm R, \bm V|t)$ is the time-dependent probability density of microstates, and $\bm a_i=d\bm v_i/dt$ is the acceleration of
each particle, which is determined by the electric and magnetic fields through the Lorenz force,
\begin{equation}
\frac{d}{dt}\left(\gamma(\bm v_i)m_i\bm v_i\right) = q_i\left(\bm E(\bm r_i; t)+\bm v_i\times \bm B(\bm r_i; t)\right).
\end{equation}

In the case of plasma, the potentials $V$ and $\bm A$ (and therefore the fields $\bm E$ and $\bm B$) include contributions from interactions
between the charged particles themselves, as well as the external fields. Then in practice, the Liouville equation is self-consistently solved together
with Maxwell's equations,

\begin{align}
\label{eq_maxwell}
\nabla \cdot \bm E  & = \frac{\rho}{\epsilon_0}, \nonumber \\
\nabla \times \bm E & = -\frac{\partial \bm B}{\partial t}, \\
\nabla \cdot \bm B  & = 0, \nonumber \\
\nabla \times \bm B & = \mu_0\left(\bm J + \epsilon_0\frac{\partial \bm E}{\partial t}\right), \nonumber
\end{align}
where $\rho$ and $\bm J$ are the charge and current densities, respectively, which are functionals of the time-dependent
probability density of microstates $P(\bm R, \bm V|t)$. The charge density is given explicitly in terms of $P(\bm R, \bm V|t)$ by
\begin{align}
\rho(\bm r; t)  & = \Big<\sum_{i=1}^N q_i \delta(\bm r_i - \bm r)\Big>_t, \nonumber \\
                & = \sum_{i=1}^N q_i \times \int d\bm R d\bm V P(\bm R, \bm V|t) \delta(\bm r_i - \bm r),
\label{eq_dens_rho}
\end{align}
while the current density is connected to $P(\bm R, \bm V|t)$ by
\begin{align}
\bm J(\bm r; t) & = \Big<\sum_{i=1}^N q_i \bm v_i \delta(\bm r_i - \bm r)\Big>_t \nonumber \\
                & = \sum_{i=1}^N q_i \times \int d\bm R d\bm V P(\bm R, \bm V|t) \bm v_i \delta(\bm r_i - \bm r).
\label{eq_dens_J}
\end{align}

\noindent
The fact that the Liouville equation needs to be solved self-consistently with the Maxwell equations does not break any of our assumptions. This is because, for a given
dynamical process of interest, if there is a self-consistent solution $V^*(\bm r; t)$ and $\bm A^*(\bm r; t)$ of Eq. \ref{eq_continuity} and \ref{eq_maxwell}, then the system
of particles can be described by the Lagrangian in Eq. \ref{eq_lagrangian} and the energy in Eq. \ref{eq_energy} under those fixed, but time-dependent potentials.

In a steady state, by definition the microstate distribution $P(\bm R, \bm V|t)$ does not depend on $t$, therefore $\rho$ and $\bm J$ are time-independent,
according to Eqs. \ref{eq_dens_rho} and Eq. \ref{eq_dens_J}. Because of Maxwell's equations, $V^*$ and $\bm A^*$ must also be time-independent. This implies that $E(\bm R, \bm V; t)=E(\bm R, \bm V)$ and
$\partial E/\partial t = 0$, so we have

\begin{equation}
\frac{dE}{dt} = \frac{\partial E}{\partial t} + \sum_{i=1}^N\left(\bm v_i\cdot \frac{\partial E}{\partial \bm r_i} + \bm a_i\cdot \frac{\partial E}{\partial \bm v_i}\right) = 0.
\end{equation}

The steady-state solutions of the Liouville equation (Eq. \ref{eq_continuity}), denoted as $P(\bm R, \bm V|\mathcal{S})$, are such that

\begin{equation}
\sum_{i=1}^N\Big(\bm v_i\cdot \frac{\partial}{\partial \bm r_i} + \bm a_i\cdot \frac{\partial}{\partial \bm v_i}\Big)P(\bm R, \bm V|\mathcal{S})= 0.
\label{eq_steady}
\end{equation}

\noindent
Because of Jeans' theorem~\cite{Dendy1995}, $P(\bm R, \bm V|\mathcal{S})$ must depend on $\bm R$ and $\bm V$ only through the integrals of motion. However, in order to
be able to define a temperature, the solution of Eq. \ref{eq_steady} can only depend on $(\bm R, \bm V)$ through the energy, that is,
\begin{equation}
P(\bm R, \bm V|\mathcal{S}) = p(E(\bm R, \bm V)).
\label{eq_ss_ensemble}
\end{equation}

In the following, we will refer to $p(E)$ as the \emph{ensemble function} It is important to make the distinction that $p(E)$, although a non-negative quantity, is not the
probability distribution of the energy in the ensemble $\mathcal{S}$, which we will denote by $P(E|\mathcal{S})$ and is given by
\begin{equation}
P(E|\mathcal{S}) = p(E)\Omega(E),
\label{eq_prob_E}
\end{equation}
with $\Omega(\varepsilon)=\int d\bm R d\bm V \delta(\varepsilon-E(\bm R, \bm V))$ the density of states.

In the next section we will consider the notion of temperature for systems in this kind of steady state.

\section{Superstatistics}
\label{sect_superstat}

The following discussion will be framed in general terms, for any system with microstates $\bm \Gamma$ in an ensemble described by
$P(\bm \Gamma|\mathcal{S})=p(H(\bm \Gamma))$, although we have in mind of course the collisionless plasma described in the previous
section, with $\bm \Gamma=(\bm R, \bm V)$ and $H(\bm \Gamma)=E(\bm R, \bm V)$.

The formalism known as superstatistics, introduced by Beck and Cohen in 2003~\cite{Beck2003}, considers a family of ensemble functions $p(E)$ of the form

\begin{equation}
p(E) = \int_0^\infty d\beta \;f(\beta)\exp(-\beta E),
\label{eq_superstat_A}
\end{equation}
where $f(\beta)$ is a non-negative weight function. From a Bayesian point of view~\cite{Sattin2006}, the correct way of writing the superstatistical ensemble is
through the marginalization of $\beta$ from the joint distribution $P(\bm \Gamma, \beta|\mathcal{S})$, that is,

\begin{align}
P(\bm \Gamma|\mathcal{S}) & = \int_0^\infty d\beta\;P(\beta|\mathcal{S}) P(\bm \Gamma|\beta) \nonumber \\
                          & = \int_0^\infty d\beta\;P(\beta|\mathcal{S}) \left[\frac{\exp(-\beta H(\bm \Gamma))}{Z(\beta)}\right].
\label{eq_superstat_B}
\end{align}
\noindent
This form of superstatistics, where the partition function appears explicitly, is called type-B superstatistics, in order to distinguish it from the form in Eq. \ref{eq_superstat_A},
which is then called type-A superstatistics. Here we define the proper probability distribution of $\beta$ as $$P(\beta|\mathcal{S}) = f(\beta)Z(\beta).$$

Although this probability distribution for $\beta$ is commonly interpreted in terms of fluctuations of a phase-space observable $\hat{\beta}(\bm \Gamma)$, this additional assumption can 
lead to inconsistencies~\cite{Davis2018} and is not strictly needed by the theory.

\section{The fundamental temperature}
\label{sect_fundamental}

In the case of a steady-state ensemble, as defined in Eq. \ref{eq_ss_ensemble}, it is possible to define an ensemble-dependent temperature function

\begin{equation}
\beta_F(E) := -\frac{d}{dE}\ln p(E),
\end{equation}
which we will call the \emph{fundamental inverse temperature}. The motivation for using this quantity can be understood starting from the canonical
equipartition theorem written in the form

\begin{equation}
\Big<\nabla \cdot \bm \omega\Big>_\beta = \beta\Big<\bm \omega\cdot \nabla H\Big>_\beta.
\label{eq_cvt_canon}
\end{equation}

Now consider $\beta$ as a random variable under a new superstatistical ensemble $\mathcal{S}$, taking expectation on both sides over the joint distribution
$P(\bm \Gamma, \beta|\mathcal{S})$. Because $\beta\big<A\big>_\beta = \big<\beta A\big>_\beta$ for any quantity $A$ we obtain
\begin{align}
\Big<\nabla \cdot \bm \omega\Big>_\mathcal{S} & = \Big< \Big<\nabla \cdot \bm{\omega}\Big>_\beta \Big>_\mathcal{S} \nonumber \\
                                              & = \Big< \beta\Big<\bm \omega\cdot \nabla H\Big>_\beta \Big>_\mathcal{S} \nonumber \\
                                              & = \Big<\beta \bm \omega\cdot \nabla H\Big>_\mathcal{S}.
\label{eq_cvt_gen}
\end{align}

On the other hand, applying the conjugate variables theorem~\cite{Davis2012} to the distribution $P(\bm \Gamma|\mathcal{S})=p(H(\bm \Gamma))$ and using
the chain rule for the gradient of $\ln P(\bm\Gamma|\mathcal{S})$, we see that~\cite{Palma2016}

\begin{align}
\Big<\nabla \cdot \bm{\omega}\Big>_\mathcal{S} & = -\Big<\bm{\omega}\cdot\nabla \ln P(\bm \Gamma|\mathcal{S})\Big>_\mathcal{S} \nonumber \\
                                               & = \Big<\beta_F(H)\bm{\omega}\cdot\nabla H\Big>_\mathcal{S}.
\label{eq_cvt}
\end{align}

\noindent
As both Eqs. \ref{eq_cvt_gen} and \ref{eq_cvt} must be valid for any choice of $\bm \omega$, we find that
\begin{equation}
\Big<\beta \bm \omega\cdot \nabla H\Big>_\mathcal{S} = \Big<\beta_F(H)\bm{\omega}\cdot\nabla H\Big>_\mathcal{S},
\end{equation}
hence
\begin{equation}
\Big<\beta G(\bm \Gamma)\Big>_\mathcal{S} = \Big<\beta_F(H) G(\bm \Gamma)\Big>_\mathcal{S}
\label{eq_property}
\end{equation}
for any function $G(\bm \Gamma)$. With the particular choice $G(\bm \Gamma) = \delta(E-H(\bm \Gamma))$ we see that

\begin{equation}
\beta_F(E) = \frac{\Big<\beta\delta(E-H(\bm \Gamma))\Big>_\mathcal{S}}{\Big<\delta(E-H(\bm \Gamma))\Big>_\mathcal{S}} = \Big<\beta\Big>_{\mathcal{S}, H=E}.
\end{equation}

In other words, the fundamental (inverse) temperature function at an energy $E$ is the mean superstatistical (inverse) temperature for all states at energy $E$. It 
also follows from Eq. \ref{eq_property} with $G=1$ that $$\Big<\beta_F\Big>_\mathcal{S} = \Big<\beta\Big>_\mathcal{S}.$$

\section{The single-particle fundamental temperature}

We are now ready to approach the problem of determining the possible velocity distributions of a particle in a steady-state plasma. At this point then, we will separate the
energy of the system into two contributions, the kinetic energy $k_1$ of a single particle (let us say particle 1 without loss of generality) and the energy $\tilde E$
of its environment, so that $$\tilde E = E - k_1 = \Phi + \sum_{i=2}^N k_i.$$

Integration of Eq. \ref{eq_ss_ensemble} over the positions $\bm R$ and the velocities $\bm{v}_2, \ldots, \bm{v}_N$ gives the probability density of $\bm v_1$ as

\begin{align}
P(\bm v_1|\mathcal{S}) & = \int d\bm R d\bm v_2\ldots d\bm v_N p(\Phi(\bm R)+K_1(\bm v_1)+ \sum_{i=2}^N K_i(\bm v_i)) \nonumber \\
                       & = \int d\tilde E\Omega_{\tilde E}(\tilde E) p(\tilde E + K_1(\bm v_1)) \nonumber \\
                       & := p_1(K_1(\bm v_1)),
\label{eq_ss_1}
\end{align}
where we have defined the density of states of the environment $\tilde E$ as
\begin{equation}
\Omega_{\tilde E}(\epsilon) = \int d\bm R d\bm v_2\ldots d\bm v_N \delta(\tilde{E}(\bm R, \bm v_2,\ldots, \bm v_N)-\epsilon).
\end{equation}

Eq. \ref{eq_ss_1} gives us the connection between the ensemble function $p(E)$ describing the full distribution of microstates and the ensemble function $p_1(k)$ that describes
the velocity distribution of a single particle. It shows that $\bm v_1$ follows a distribution with ensemble function
\begin{equation}
p_1(k) = \int d\tilde E\Omega_{\tilde E}(\tilde E)p(\tilde E + k),
\label{eq_p1}
\end{equation}
and then the probability of observing a kinetic energy $k_1$ in the ensemble $\mathcal{S}$ is given simply by
\begin{equation}
P(K_1=k|\mathcal{S})=p_1(k)\Omega_{K_1}(k),
\end{equation}
in agreement with Eq. \ref{eq_prob_E}. Similarly, the joint distribution of energies of the subsystem $k$ and the environment $\tilde E$ is given by
\begin{equation}
P(\tilde E, k|\mathcal{S})=p(\tilde E + k)\Omega_{K_1}(k)\Omega_{\tilde E}(\tilde E).
\label{eq_p_joint}
\end{equation}

From the ensemble function $p_1(k)$ in Eq. \ref{eq_p1} we will compute directly the single-particle fundamental temperature as

\begin{align}
\beta^{(1)}_F(k) & = -\frac{d}{dk}\ln p_1(k) \nonumber \\
                 & = \int d\tilde E\left[\frac{\Omega_{\tilde E}(\tilde E)p(E)}{p_1(k)}\right]\left\{-\frac{d}{dE}\ln p(E)\right\},
\label{eq_p_cond}
\end{align}
which is the expectation of $\beta_F(E)$ (inside curly braces) under a particular distribution for $E$ (inside square brackets). Now we use the product
rule of probability theory~\cite{Jaynes2003} to show that
\begin{align}
P(E|\mathcal{S}, K_1=k) & = \frac{P(E, K_1=k|\mathcal{S})}{P(K_1=k|\mathcal{S})} \nonumber \\
                        & = \frac{p(E)\Omega_{\tilde E}(\tilde E)\Omega_{K_1}(k)}{p_1(k)\Omega_{K_1}(k)} \nonumber \\
                        & = \frac{p(E)\Omega_{\tilde E}(\tilde E)}{p_1(k)}.
\end{align}

\noindent
Replacing in Eq. \ref{eq_p_cond} we can finally obtain the single-particle fundamental (inverse) temperature as
\begin{align}
\beta^{(1)}_F(k) & = \int dE P(E|\mathcal{S}, K_1=k) \beta_F(E) \nonumber \\
                 & = \Big<\beta_F(E)\Big>_{\mathcal{S},K_1=k}.
\end{align}

This is an important result. It shows that the fundamental (inverse) temperature for the velocity of a single particle at energy $k$ is given by the conditional
expectation of the fundamental (inverse) temperature for the complete system, when the kinetic energy $K_1(\bm v_1)$ of the particle is fixed at $k$.

Here we can distinguish two cases:

\begin{enumerate}

\item [(a)] If the effect of fixing the kinetic energy $K_1$ on the whole system is negligible, then $\beta^{(1)}_F(k)$ will be a constant $\beta_0$, independent 
of $k$, and the particle will then be described by the canonical ensemble
\begin{equation}
P(\bm v_1|\beta_0) \propto \exp\left(-\beta_0 K_1(\bm v_1)\right)
\end{equation}
at inverse temperature $\beta_0=\big<\beta_F(E)\big>_{\mathcal{S}}$. Note that because $K_1(\bm v_1)=\gamma(v_1)m_1c^2$, this coincides with the Jüttner
distribution~\cite{Nakamura2009} for relativistic speeds, and reduces to the Maxwell-Boltzmann distribution for non-relativistic particles.

\item [(b)] If the particles in the system are highly correlated (as in the case of plasma), then fixing $K_1 = k$ does influence the states of the system, and therefore
\begin{equation}
\big<\beta_F(E)\big>_{\mathcal{S},K_1=k} \neq \big<\beta_F(E)\big>_{\mathcal{S}}.
\label{eq_betaf_noncanon}
\end{equation}

\noindent
In this case, $\beta^{(1)}_F(k)$ will depend on $k$, producing a non-canonical ensemble.

\end{enumerate}

In the next section we will show the conditions under which $P(\bm v_1|\mathcal{S})$ of Eq. \ref{eq_ss_1} can be described by Superstatistics,
and obtain an explicit expression for the superstatistical (inverse) temperature distribution $P(\beta|\mathcal{S})$.

\section{Derivation of the superstatistical formalism}
\label{sect_derivation}

In order to simplify the description of the single-particle velocity distribution given by Eqs. \ref{eq_ss_1} and \ref{eq_p1}, we will make the reasonable assumption
that the energy $\tilde E$ of the environment is much larger than the kinetic energy of the particle, $\tilde E \gg k$. This is to be expected in plasmas because then
the kinetic energies dominate over the potential energy~\cite{Bellan2006}, so $\Phi$ will not be able to cancel $K'=K-k_1$, even when the system is bound ($\Phi < 0$).
Under this assumption we can use the definition of $\beta_F(E)$ to approximate $p(\tilde E + k)$ in Eq. \ref{eq_p1} as
\begin{align*}
p(\tilde E + k) & = \exp\left(\ln p(\tilde E + k)\right) \nonumber \\
                & \approx \exp\left(\ln p(\tilde E) + k\frac{d}{d \tilde E}\ln p(\tilde E)\right) \nonumber \\
                & = p(\tilde E)\exp(-\beta_F(\tilde E)k),
\end{align*}

\noindent
and hence obtain
\begin{equation}
p_1(k) \approx \int d\tilde E\Omega_{\tilde E}(\tilde E) p(\tilde E)\exp(-\beta_F(\tilde E)k).
\end{equation}

Now, in the case when $\beta_F(E) > 0$ we can introduce the function
\begin{equation}
f(\beta) = \int d\tilde E\Omega_{\tilde E}(\tilde E) p(\tilde E)\delta(\beta - \beta_F(\tilde E))
\label{eq_fbeta}
\end{equation}
and use it to rewrite $p_1(k)$ as
\begin{equation}
p_1(k) = \int_0^\infty d\beta f(\beta)\exp(-\beta k).
\end{equation}

\noindent
This is precisely type-A superstatistics, while the corresponding expression in type-B superstatistics is obtained by including the single-particle partition function $Z_1$ as
\begin{align}
p_1(k)  & = \int_0^\infty d\beta f(\beta)Z_1(\beta)\left[\frac{\exp(-\beta k)}{Z_1(\beta)}\right] \nonumber \\
        & = \int_0^\infty d\beta P(\beta|\mathcal{S})\left[\frac{\exp(-\beta k)}{Z_1(\beta)}\right],
\label{eq_p1_final}
\end{align}
where
\begin{align*}
Z_1(\beta) & = \int_0^\infty dk \Omega_{K_1}(k)\exp(-\beta k) \nonumber \\
           & = Z_0\frac{\exp(\beta m_1c^2)}{\beta m_1c^2}\mathcal{K}_2(\beta m_1c^2),
\end{align*}
and $\mathcal{K}_2(x)$ is the modified Bessel function of the second kind~\cite{Greiner2012}. We can give an explicit expression for $P(\beta|\mathcal{S})$ as

\begin{widetext}
\begin{align}
P(\beta|\mathcal{S}) & = \int d\tilde E\Omega_{\tilde E}(\tilde E) p(\tilde E)\delta(\beta - \beta_F(\tilde E))Z_1(\beta) \nonumber \\
                 & = \int d\tilde E\Omega_{\tilde E}(\tilde E) p(\tilde E)\delta(\beta - \beta_F(\tilde E))\times\Big(\int dk\Omega_{K_1}(k)\exp(-\beta k)\Big) \nonumber \\
                 & = \int dk d\tilde E\Omega_{K_1}(k)\Omega_{\tilde E}(\tilde E) p(\tilde E)\exp(-\beta k)\delta(\beta - \beta_F(\tilde E)) \nonumber \\
                 & = \int dk d\tilde E\Omega_{K_1}(k)\Omega_{\tilde E}(\tilde E) p(\tilde E)\exp(-\beta_F(\tilde E)k)\delta(\beta - \beta_F(\tilde E)) \nonumber \\
                 & \approx \int dk d\tilde E\left[\Omega_{K_1}(k)\Omega_{\tilde E}(\tilde E) p(E)\right]\delta(\beta - \beta_F(\tilde E)) = \int d\tilde E P(\tilde E|\mathcal{S})\delta(\beta-\beta_F(\tilde E)),
\label{eq_pbeta}
\end{align}
\end{widetext}
where in the last equality we have used Eq. \ref{eq_p_joint}, together with the marginalization rule as
\begin{displaymath}
\int dk P(k, \tilde E|\mathcal{S}) = P(\tilde E|\mathcal{S}).
\end{displaymath}
\noindent
Eq. \ref{eq_pbeta} then tells us that $P(\beta|\mathcal{S}) \approx P(\beta_F(\tilde E)=\beta | \mathcal{S})$ and therefore
\begin{equation}
P(\bm v_1|\mathcal{S}) \approx \int_0^\infty\hspace{-7pt} d\beta P(\beta_F(\tilde E) = \beta|\mathcal{S})\left[\frac{\exp(-\beta K_1(\bm v_1))}{Z_1(\beta)}\right].
\label{eq_final}
\end{equation}

This approximation allows us to understand the fluctuating temperature of the particle \emph{not in terms of its kinetic energy, but as the fundamental
temperature of its environment}. A direct consequence of this is the fact that a non-fluctuating environment will always induce an equilibrium distribution for $\bm v_1$,
because then
\begin{displaymath}
P(\beta_F(\tilde E) = \beta|\mathcal{S}) = P(\beta|\mathcal{S}) = \delta(\beta - \beta_F(E_0)),
\end{displaymath}
where $E_0$ is the constant energy of the environment, and in this case superstatistics (Eq. \ref{eq_superstat_B}) reduces to the canonical ensemble.

\section{Concluding Remarks}
\label{sect_conclusions}

In summary, we have shown that it is possible to determine the velocity distribution function of a particle in a plasma given the original ensemble function of the
complete system (or, equivalently, its fundamental temperature function), and the probability distribution of the environment energy $\tilde E$. This distribution belongs
to the superstatistics family, and its parameter $\beta$ follows the same distribution as $\beta_F(\tilde E)$. 

Our interpretation of the superstatistical $\beta$ for the single particle velocity distribution is then connected to the energy fluctuations of the environment (and not of the system 
itself), and in absence of them, the particle follows a canonical (Jüttner or Maxwell-Boltzmann) distribution. If the particle is highly correlated with its environment (as is the case 
with plasmas), then in general the particle velocities are non-Maxwellian, because the fundamental temperature will be energy-dependent, as per Eq. \ref{eq_betaf_noncanon}. In particular, 
it follows from Eq. \ref{eq_final} that a sufficient condition for Kappa/Tsallis statistics is for $\beta_F(\tilde E)$ to follow a gamma distribution.

\section{Acknowledgements}
This work was supported by FONDECYT 1171127 and Anillo ACT-172101 grants.

\bibliography{plasma_nonextens}

\begin{thebibliography}{21}
\expandafter\ifx\csname natexlab\endcsname\relax\def\natexlab#1{#1}\fi
\expandafter\ifx\csname bibnamefont\endcsname\relax
  \def\bibnamefont#1{#1}\fi
\expandafter\ifx\csname bibfnamefont\endcsname\relax
  \def\bibfnamefont#1{#1}\fi
\expandafter\ifx\csname citenamefont\endcsname\relax
  \def\citenamefont#1{#1}\fi
\expandafter\ifx\csname url\endcsname\relax
  \def\url#1{\texttt{#1}}\fi
\expandafter\ifx\csname urlprefix\endcsname\relax\def\urlprefix{URL }\fi
\providecommand{\bibinfo}[2]{#2}
\providecommand{\eprint}[2][]{\url{#2}}

\bibitem[{\citenamefont{Lima et~al.}(2000)\citenamefont{Lima, Silva, and
  Santos}}]{Lima2000}
\bibinfo{author}{\bibfnamefont{J.}~\bibnamefont{Lima}},
  \bibinfo{author}{\bibfnamefont{R.}~\bibnamefont{Silva}}, \bibnamefont{and}
  \bibinfo{author}{\bibfnamefont{J.}~\bibnamefont{Santos}},
  \bibinfo{journal}{Physical Review E} \textbf{\bibinfo{volume}{61}},
  \bibinfo{pages}{3260} (\bibinfo{year}{2000}).

\bibitem[{\citenamefont{Leubner}(2002)}]{Leubner2002}
\bibinfo{author}{\bibfnamefont{M.~P.} \bibnamefont{Leubner}},
  \bibinfo{journal}{Astrophysics and Space Science}
  \textbf{\bibinfo{volume}{282}}, \bibinfo{pages}{573} (\bibinfo{year}{2002}).

\bibitem[{\citenamefont{Livadiotis}(2017)}]{Livadiotis2017}
\bibinfo{author}{\bibfnamefont{G.}~\bibnamefont{Livadiotis}},
  \emph{\bibinfo{title}{Kappa distributions: Theory and applications in
  plasmas}} (\bibinfo{publisher}{Elsevier}, \bibinfo{year}{2017}).

\bibitem[{\citenamefont{Gougam and Tribeche}(2011)}]{Gougam2011}
\bibinfo{author}{\bibfnamefont{L.~A.} \bibnamefont{Gougam}} \bibnamefont{and}
  \bibinfo{author}{\bibfnamefont{M.}~\bibnamefont{Tribeche}},
  \bibinfo{journal}{Physics of Plasmas} \textbf{\bibinfo{volume}{18}},
  \bibinfo{pages}{62102} (\bibinfo{year}{2011}).

\bibitem[{\citenamefont{Viñas et~al.}(2015)\citenamefont{Viñas, Moya,
  Navarro, Valdivia, Araneda, and Muñoz}}]{Vinas2015}
\bibinfo{author}{\bibfnamefont{A.~F.} \bibnamefont{Viñas}},
  \bibinfo{author}{\bibfnamefont{P.~S.} \bibnamefont{Moya}},
  \bibinfo{author}{\bibfnamefont{R.~E.} \bibnamefont{Navarro}},
  \bibinfo{author}{\bibfnamefont{J.~A.} \bibnamefont{Valdivia}},
  \bibinfo{author}{\bibfnamefont{J.~A.} \bibnamefont{Araneda}},
  \bibnamefont{and} \bibinfo{author}{\bibfnamefont{V.}~\bibnamefont{Muñoz}},
  \bibinfo{journal}{J. Geophys. Res: Space Phys.}
  \textbf{\bibinfo{volume}{120}}, \bibinfo{pages}{3307} (\bibinfo{year}{2015}).

\bibitem[{\citenamefont{Livadiotis}(2014)}]{Livadiotis2014}
\bibinfo{author}{\bibfnamefont{G.}~\bibnamefont{Livadiotis}},
  \bibinfo{journal}{Entropy} \textbf{\bibinfo{volume}{16}},
  \bibinfo{pages}{4290} (\bibinfo{year}{2014}).

\bibitem[{\citenamefont{Livadiotis}(2016)}]{Livadiotis2016}
\bibinfo{author}{\bibfnamefont{G.}~\bibnamefont{Livadiotis}},
  \bibinfo{journal}{EPL (Europhysics Letters)} \textbf{\bibinfo{volume}{113}},
  \bibinfo{pages}{10003} (\bibinfo{year}{2016}).

\bibitem[{\citenamefont{Tsallis}(2009)}]{Tsallis2009}
\bibinfo{author}{\bibfnamefont{C.}~\bibnamefont{Tsallis}},
  \emph{\bibinfo{title}{Introduction to nonextensive statistical mechanics:
  approaching a complex world}} (\bibinfo{publisher}{Springer Science \&
  Business Media}, \bibinfo{year}{2009}).

\bibitem[{\citenamefont{Beck}(2004)}]{Beck2004}
\bibinfo{author}{\bibfnamefont{C.}~\bibnamefont{Beck}},
  \bibinfo{journal}{Continuum Mechanics and Thermodynamics}
  \textbf{\bibinfo{volume}{16}}, \bibinfo{pages}{293} (\bibinfo{year}{2004}).

\bibitem[{\citenamefont{Ourabah et~al.}(2015)\citenamefont{Ourabah, Gougam, and
  Tribeche}}]{Ourabah2015}
\bibinfo{author}{\bibfnamefont{K.}~\bibnamefont{Ourabah}},
  \bibinfo{author}{\bibfnamefont{L.~A.} \bibnamefont{Gougam}},
  \bibnamefont{and} \bibinfo{author}{\bibfnamefont{M.}~\bibnamefont{Tribeche}},
  \bibinfo{journal}{Physical Review E} \textbf{\bibinfo{volume}{91}},
  \bibinfo{pages}{12133} (\bibinfo{year}{2015}).

\bibitem[{\citenamefont{Ichimaru}(1992)}]{Ichimaru1992}
\bibinfo{author}{\bibfnamefont{S.}~\bibnamefont{Ichimaru}},
  \emph{\bibinfo{title}{Statistical Plasma Physics: Volume 1}}
  (\bibinfo{year}{1992}).

\bibitem[{\citenamefont{Dendy}(1995)}]{Dendy1995}
\bibinfo{author}{\bibfnamefont{R.~O.} \bibnamefont{Dendy}},
  \emph{\bibinfo{title}{Plasma Physics: An Introductory Course}}
  (\bibinfo{publisher}{Cambridge University Press}, \bibinfo{year}{1995}).

\bibitem[{\citenamefont{Beck and Cohen}(2003)}]{Beck2003}
\bibinfo{author}{\bibfnamefont{C.}~\bibnamefont{Beck}} \bibnamefont{and}
  \bibinfo{author}{\bibfnamefont{E.}~\bibnamefont{Cohen}},
  \bibinfo{journal}{Physica A} \textbf{\bibinfo{volume}{322}},
  \bibinfo{pages}{267} (\bibinfo{year}{2003}).

\bibitem[{\citenamefont{Sattin}(2006)}]{Sattin2006}
\bibinfo{author}{\bibfnamefont{F.}~\bibnamefont{Sattin}},
  \bibinfo{journal}{European Physical Journal B} \textbf{\bibinfo{volume}{49}},
  \bibinfo{pages}{219} (\bibinfo{year}{2006}).

\bibitem[{\citenamefont{Davis and Gutiérrez}(2018)}]{Davis2018}
\bibinfo{author}{\bibfnamefont{S.}~\bibnamefont{Davis}} \bibnamefont{and}
  \bibinfo{author}{\bibfnamefont{G.}~\bibnamefont{Gutiérrez}},
  \bibinfo{journal}{Physica A} \textbf{\bibinfo{volume}{505}},
  \bibinfo{pages}{864} (\bibinfo{year}{2018}).

\bibitem[{\citenamefont{Davis and Guti\'errez}(2012)}]{Davis2012}
\bibinfo{author}{\bibfnamefont{S.}~\bibnamefont{Davis}} \bibnamefont{and}
  \bibinfo{author}{\bibfnamefont{G.}~\bibnamefont{Guti\'errez}},
  \bibinfo{journal}{Phys. Rev. E} \textbf{\bibinfo{volume}{86}},
  \bibinfo{pages}{051136} (\bibinfo{year}{2012}).

\bibitem[{\citenamefont{Palma et~al.}(2016)\citenamefont{Palma, Gutiérrez, and
  Davis}}]{Palma2016}
\bibinfo{author}{\bibfnamefont{G.}~\bibnamefont{Palma}},
  \bibinfo{author}{\bibfnamefont{G.}~\bibnamefont{Gutiérrez}},
  \bibnamefont{and} \bibinfo{author}{\bibfnamefont{S.}~\bibnamefont{Davis}},
  \bibinfo{journal}{Phys. Rev. E} \textbf{\bibinfo{volume}{94}},
  \bibinfo{pages}{062113} (\bibinfo{year}{2016}).

\bibitem[{\citenamefont{Jaynes}(2003)}]{Jaynes2003}
\bibinfo{author}{\bibfnamefont{E.~T.} \bibnamefont{Jaynes}},
  \emph{\bibinfo{title}{Probability Theory: The Logic of Science}}
  (\bibinfo{publisher}{Cambridge University Press}, \bibinfo{year}{2003}).

\bibitem[{\citenamefont{Nakamura}(2009)}]{Nakamura2009}
\bibinfo{author}{\bibfnamefont{T.~K.} \bibnamefont{Nakamura}},
  \bibinfo{journal}{EPL (Europhysics Letters)} \textbf{\bibinfo{volume}{88}},
  \bibinfo{pages}{40009} (\bibinfo{year}{2009}).

\bibitem[{\citenamefont{Bellan}(2006)}]{Bellan2006}
\bibinfo{author}{\bibfnamefont{P.~M.} \bibnamefont{Bellan}},
  \emph{\bibinfo{title}{Fundamentals of Plasma Physics}}
  (\bibinfo{publisher}{Cambridge University}, \bibinfo{year}{2006}).

\bibitem[{\citenamefont{Greiner et~al.}(2012)\citenamefont{Greiner, Neise, and
  St{\"o}cker}}]{Greiner2012}
\bibinfo{author}{\bibfnamefont{W.}~\bibnamefont{Greiner}},
  \bibinfo{author}{\bibfnamefont{L.}~\bibnamefont{Neise}}, \bibnamefont{and}
  \bibinfo{author}{\bibfnamefont{H.}~\bibnamefont{St{\"o}cker}},
  \emph{\bibinfo{title}{Thermodynamics and statistical mechanics}}
  (\bibinfo{publisher}{Springer Science \& Business Media},
  \bibinfo{year}{2012}).

\end{thebibliography}
\bibliographystyle{apsrev}

\end{document}